\documentclass[12pt]{article}

\usepackage[utf8]{inputenc}
\usepackage{scicite}
\usepackage{soul,color}
\usepackage{graphicx}
\usepackage{times}
\usepackage{caption}
\usepackage{subcaption}
\usepackage{booktabs}
\usepackage{float}
\usepackage{hyperref}
\usepackage{blindtext}

\newenvironment{sciabstract}{%
\begin{quote} \bf}
{\end{quote}}

\title{Mobile phone location data reveal the effect and geographic variation of social distancing on the spread of the COVID-19 epidemic} 

\author
{
	Song Gao$^{1\ast\dagger}$, Jinmeng Rao$^{1\ast}$, Yuhao Kang$^{1\ast}$, Yunlei Liang$^{1\ast}$, Jake Kruse$^{1}$,\\Doerte Doepfer$^{2}$,  Ajay K. Sethi$^{3}$, Juan Francisco Mandujano Reyes$^{2}$,  \\
Jonathan Patz$^{3}$, Brian S. Yandell$^{4}$
\\
\normalsize{$^{1}$GeoDS Lab, Department of Geography, University of Wisconsin-Madison, WI 53706, USA}\\
\normalsize{$^{2}$School Of Veterinary Medicine, University of Wisconsin-Madison, WI 53706, USA}\\
\normalsize{$^{3}$School of Medicine and Public Health, University of Wisconsin-Madison, WI 53706, USA}\\
\normalsize{$^{4}$Statistics and American Family Insurance Data Science Institute} \\\normalsize{University of Wisconsin-Madison, WI 53706, USA}\\
\\
\normalsize{$^\ast$ These authors contributed equally to this work.}
\\
\normalsize{$^\dagger$To whom correspondence should be addressed; E-mail: song.gao@wisc.edu.}
}

\date{}

\begin{document} 
\baselineskip24pt
\maketitle 

\begin{sciabstract} 
The emergence of SARS-CoV-2 and the coronavirus infectious disease (COVID-19) has become a pandemic. Social (physical) distancing is a key non-pharmacologic control measure to reduce the transmission rate of SARS-COV-2, but high-level adherence is needed. Using daily travel distance and stay-at-home time derived from large-scale anonymous mobile phone location data provided by Descartes Labs and SafeGraph, we quantify the degree to which social distancing mandates have been followed in the U.S. and its effect on growth of COVID-19 cases. The correlation between the COVID-19 growth rate and travel distance decay rate and dwell time at home change rate was -0.586 (95\% CI: -0.742$\sim$-0.370) and 0.526 (95\% CI: 0.293$\sim$0.700), respectively. Increases in state-specific doubling time of total cases ranged from 1.04$\sim$6.86 days to 3.66$\sim$30.29 days after social distancing orders were put in place, consistent with mechanistic epidemic prediction models. Social distancing mandates reduce the spread of COVID-19 when they are followed. 
\end{sciabstract}

\section*{Introduction}
\begin{figure}[H]
	\centering
	\includegraphics[scale = 0.5]{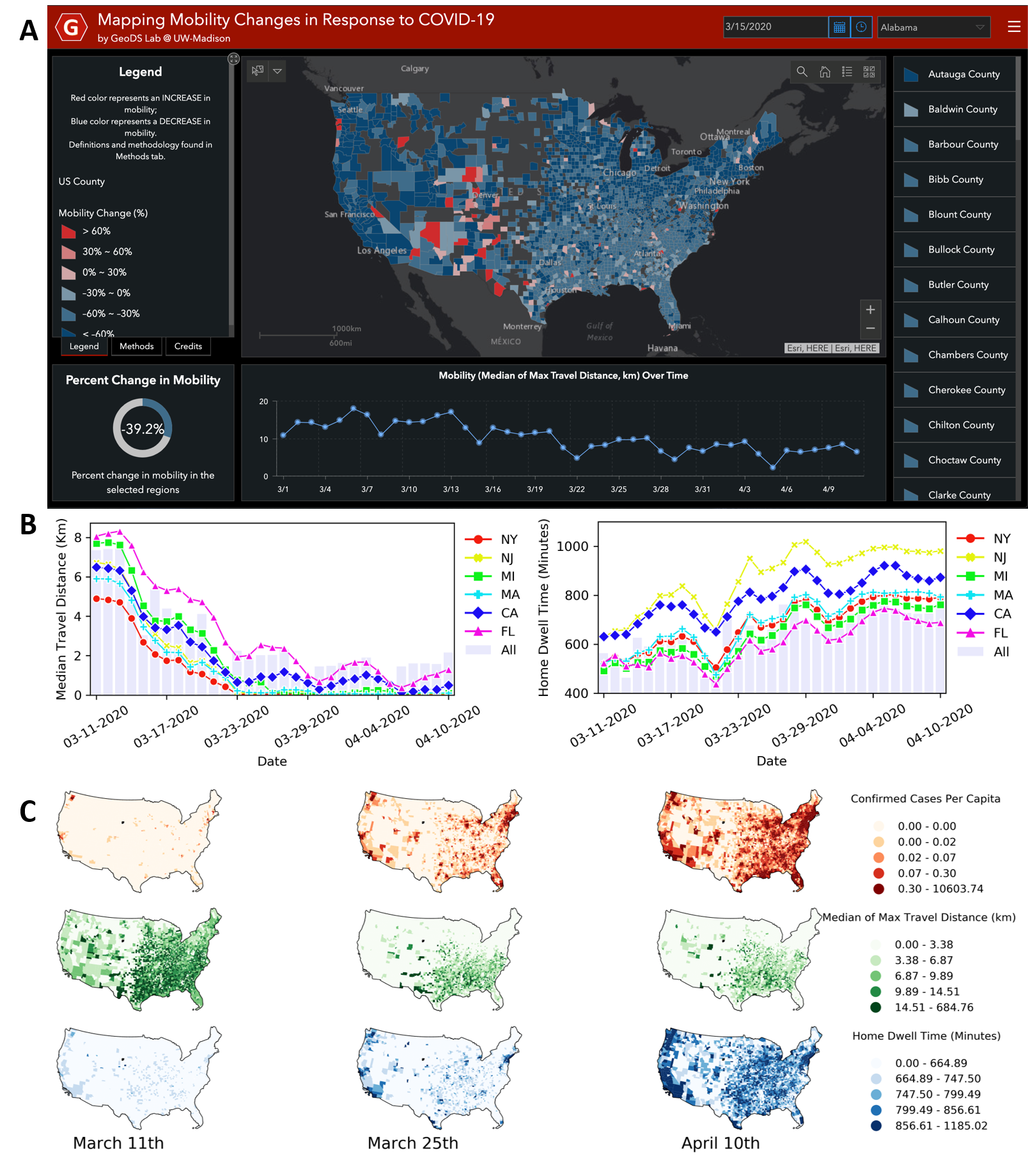}
	\caption{A. The web mapping platform for tracking human mobility changes at the county level in the United States (showing the spatial pattern on March 15, 2020 and available at \url{https://geods.geography.wisc.edu/covid19/physical-distancing/}). B. The temporal changes of the median of individual maximum travel distance (left) and the median of home dwell time (right) in the most infected U.S. states from March 11 to April 10, 2020. C. The comparison among confirmed cases per capta, median of individual maximum travel distance, median of home dwell time on March 11, March 25, and April 10, 2020.}
	\label{fig:geods_webportal}
\end{figure}

The coronavirus disease (COVID-19) pandemic is a global threat with escalating health, economic and social challenges. As of April 11, 2020, there had been 492,416 total confirmed cases and 18,559 total deaths in the U.S. according to the reports of the Centers for Disease Control and Prevention (CDC) \cite{CDC-transmission}. 
People are still witnessing widespread community transmission of the COVID-19 all over the world. Presently, there is neither a vaccine nor pharmacologic agent found to reduce the transmission of severe acute respiratory syndrome coronavirus-2 (SARS-CoV-2), the virus that causes COVID-19. Thus, the effects of non-pharmacological epidemic control and intervention measures including travel restrictions, closures of schools and nonessential business services, wearing of face masks, testing, isolation and timely quarantine on delaying the COVID-19 spread have been largely investigated and reported \cite{pan2020association,hartley2020jama,lai2020effect,chinazzi2020effect,tian2020impact}.  To mitigate and ultimately contain the COVID-19 epidemic, one of the important (non-pharmacological) control measures to reduce the transmission rate of SARS-COV-2 in the population is social (physical) distancing. An interactive web-based mapping platform (in Fig. \ref{fig:geods_webportal}A) that provides timely quantitative information on how people in different counties and states reacted to the social distancing guidelines was developed \cite{gao2020mapping}. It integrates geographic information systems (GIS) and daily updated human mobility statistical patterns derived from large-scale anonymized and aggregated smartphone location big data at the county-level in the U.S. \cite{zhou2020covid,descarteslabs2020mobility,prestby2019understanding,liang2020calibrating}. The primary goal of the online platform is to increase risk awareness among the public, support governmental decision-making, and help enhance U.S. community responses to the COVID-19 pandemic. 

It is worth noting that reduced mobility does not necessarily ensure the social (physical) distancing in practice following the CDC's definition: ``Stay at least 6 feet (2 meters) from other people" \cite{CDC-socialdistancing}. Due to the mobile phone GPS horizontal error and uncertainty \cite{gao20171}, such physical distancing patterns cannot be directly identified from the used aggregated mobility data; it requires other wearable sensors or bluetooth trackers, which raise issues of personal data privacy and ethical concerns \cite{buckee2020aggregated}. Because COVID-19 is twice as contagious and far more deadly than seasonal flu, social (physical) distancing is critical in our fight to save lives and prevent suffering. However, so far, to what degree such guidelines have been followed from place to place before and after shelter-in-place orders across the U.S. and the quantitative effect on flattening the curve were unknown. 

To this end, we employed two social distancing metrics: the median of individual maximum travel distance and the home dwell time derived from large-scale mobile phone location data (in Fig. \ref{fig:geods_webportal} B-C)  provided by Descartes Labs and SafeGraph to assess the effectiveness of stay-at-home policies on curbing the spread of the COVID-19 epidemic. For each state, we examined these measures against the growth rate of SARS-CoV-2 transmission. 

\section*{Findings}

\subsection*{The relationship between the mobility changes and the growth of the infected population}

\begin{figure}[H]
\centering
 \includegraphics[width=\textwidth]{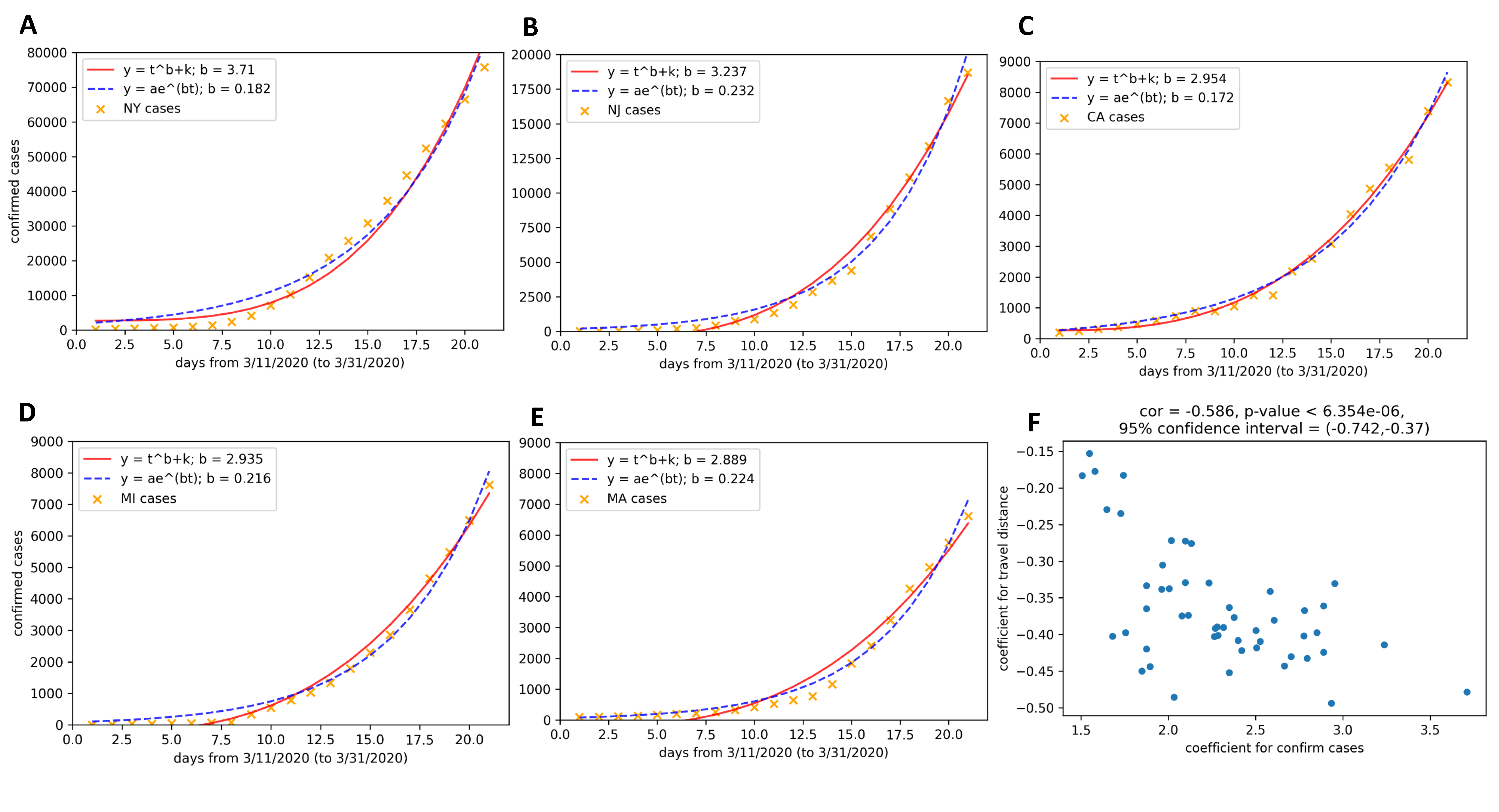}
\caption{The curve fitting results of total number of infected people for the top five states with the largest coefficients. A: New York; B: New Jersey; C: California; D: Michigan; E: Massachusetts; F: The state-level correlation between the growth coefficients of confirmed cases and the travel distance decay coefficients.}
\label{fig:curve_fitting}
\end{figure}

By fitting the curves for the state-specific confirmed cases from March 11 to April 10, 2020 using a scaling law formula \cite{maier2020effective}, we were able to identify the top five states with the largest growth rates of confirmed cases: New York, New Jersey, California, Michigan, and Massachusetts. Our fitting results corresponded to the up-to-date COVID-19 situation so far (Tables S1 and S2).  
Fig. \ref{fig:curve_fitting}A-E show the reported cases and the fitting curves in these five states using the formulas $y_c = t^b + k$ and  $y_c = ae^{bt}$, where $y_c$ is the total number of confirmed cases in each state, $t$ is the number of days from the declared date of the pandemic: March 11, 2020, and $a,b,k$ are the parameters we need to estimate (in supplementary materials). Meanwhile, we used linear regression to detect the travel distance decreasing rate over time. 
The Pearson's correlation coefficient between the cases growth rate and the distance decay rate was -0.586 (95\% CI: [-0.742, -0.370], p-value $<$0.00001). Fig. \ref{fig:curve_fitting}F shows the state-level correlation between the growth coefficients of confirmed cases and the travel distance decay coefficients across the nation. The moderate negative relationship indicated that in the states where the confirmed cases were growing faster, people responded more actively and quicker by reducing their daily travel distance.

\subsection*{The relationship between the stay-at-home duration changes and the growth of the infected population} 

We also fitted the curve for the home dwell time changes for each state using the scaling and linear models, and calculated the correlation between the home dwell time increasing rate and the growth rate of the total number of infected people. The two change rates have a positive correlation of 0.526 (95\% CI: [0.293, 0.700], p-value $<$ 0.0001), which means that in areas with higher cases growth rates, people responded better and stayed at home for longer time. 

The results of two above-mentioned association analyses both showed that there existed dramatic mobility reduction in response to the fast growth of the COVID-19 cases and people in most states reacted to the social distancing guidelines by reducing daily travel distance and increasing stay-at-home time. In return, the overall trend of reducing growth rates of cases was found later across different states although the geographic variation still existed. 

\subsection*{The effect of social distancing on delaying the epidemic doubling time} 
\begin{figure}[H]
  \centering
  \includegraphics[width=0.76\linewidth]{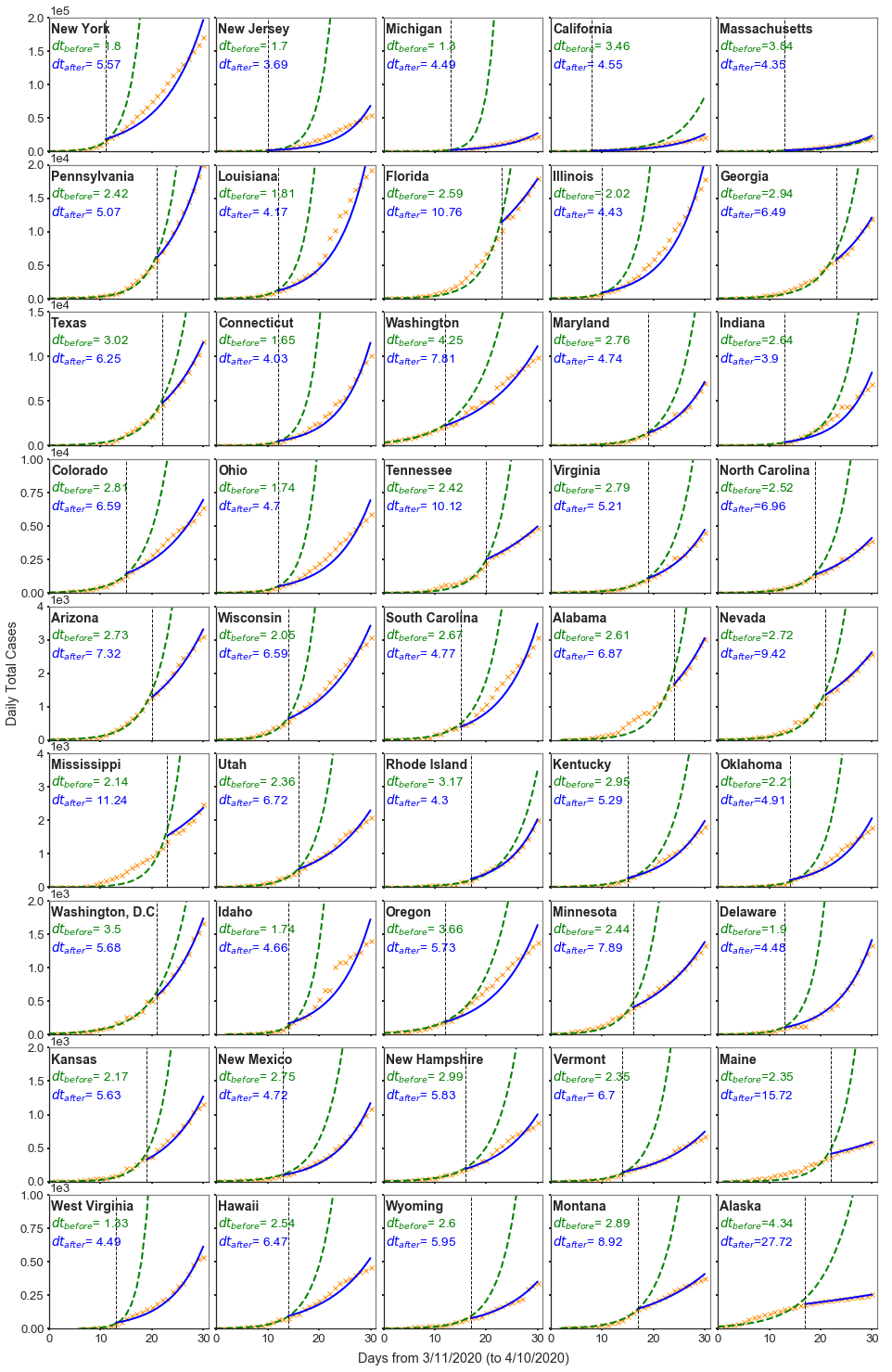}
  \caption{The curve fitting results using the exponential growth model for each state. The green dashed line and the blue line represent the fitted curves on the data before and after the stay-at-home orders in each state, respectively; the vertical black dashed line indicates the effective date of the stay-at-home orders in each state. $dt_{before}$ and $dt_{after}$ represent the median doubling time before and after the order in each state.} 
  \label{fig:dt_exp_fit}
\end{figure}

Specifically, we investigated how the social distancing guidelines and stay-at-home orders (Table S5) affected the epidemic doubling time of confirmed cases (from March 11 to April 10) in each state. We used mathematical curve fitting models and mechanistic epidemic prediction models using Bayesian parametric estimation of the serial interval distribution of successive cases to cross validate the conclusion \cite{cori2013new,thompson2019improved}. The fitted curves by an exponential model and a power-law model are shown in Fig. \ref{fig:dt_exp_fit} and Fig. S2. For the exponential model (Table S10), before the statewide stay-at-home orders, initial estimates of the growth rates of the number of infected people for the outbreak in each state were 0.17$\sim$0.70 per day with a doubling time of 1.30$\sim$4.34 days (median: 2.59 days; IQR: 0.752). A similar result was found by fitting the power-law model (Table S11), in which initial estimates of the growth rates before the orders in each state were 0.12$\sim$0.71 per day with a doubling time of 1.30$\sim$6.18 days (median: 2.71 days; IQR: 0.915). The finding aligned well with the doubling time of 2.3$\sim$3.3 days in the early outbreak epicenter Wuhan, China \cite{sanche2020high}. After the orders, the estimates of the growth rate in each state by the exponential model were reduced to 0.03$\sim$0.21 per day with a doubling time increased to 3.69$\sim$27.72 days (median: 5.68 days; IQR: 2.203). Similarly, the estimates of the growth rate in each state by the power-law model were reduced to 0.02$\sim$0.17 per day with a doubling time increased to 4.31$\sim$29.77 days (median: 6.27 days; IQR: 2.457). The finding also aligned well with the result from the observed epidemiological data (Table 1), in which the empirical growth rate in each state was 0.11$\sim$0.95 per day with a doubling time of 1.04$\sim$6.86 days (median: 2.69 days; IQR: 1.011) before the statewide stay-at-home orders, and reduced to 0.02$\sim$0.21 per day with a doubling time increased to 3.66$\sim$30.29 days (median: 5.98 days; IQR: 2.345) after the orders. The exponential equation approach was particularly suitable during the early outbreak phase and our curve fitting results matched the outcomes of mechanistic epidemic prediction models (Fig. S5 and S6 in supplementary materials), such as the models reported by 
\cite{cori2013new,thompson2019improved}. These models used confirmed cases and the serial interval, that is the days between two successive infected cases.

\begin{figure}[H]
  \centering
  \includegraphics[width=0.6\linewidth]{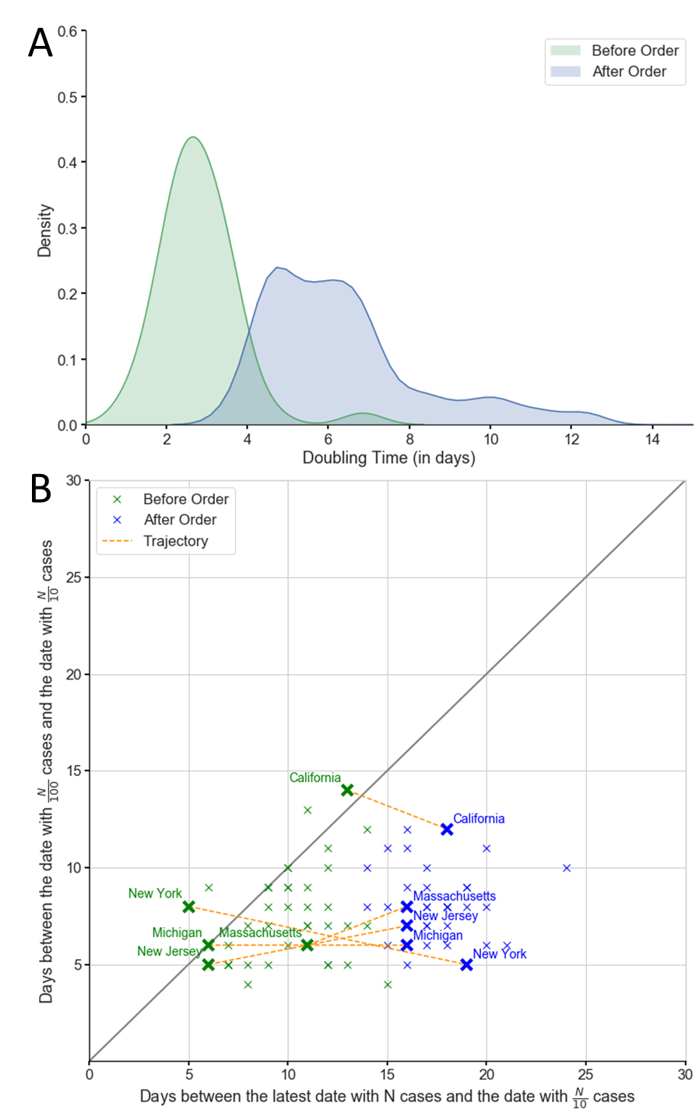}
  \caption{A. The overall changes of the doubling time nationwide before and after the stay-at-home orders using the observed epidemiological data. The median doubling time was 1.04$\sim$6.86 days (IQR: 1.011) across states before the order and increased to 3.66$\sim$30.29 days (IQR: 2.345) after the order. B. The Ten-Hundred Plot showing how fast COVID-19 spreads before and after stay-at-home orders in each state. The Lower-Right region represents sub-exponential growth, The Diagonal represents exponential growth, and the Upper-Left region represents super-exponential growth. The top five states with the most confirmed cases are labeled and their growth changes are visualized as trajectories.}
  \label{fig:overall_doublingtime_dist}
\end{figure}

Besides, we investigated the overall probability density distribution of the doubling time nationwide before and after the orders using the state-level median doubling time (Fig. \ref{fig:overall_doublingtime_dist}A, S3 and S4). The doubling time nationwide has increased from mainly 1$\sim$6 days to 3$\sim$14 days after the stay-at-home orders. The results about the doubling time all confirmed the effectiveness of social distancing in slowing down the COVID-19 transmission and in flattening the curve. The Ten-Hundred plot (Fig. \ref{fig:overall_doublingtime_dist}B) \cite{tenhundredplot} also shows that the case growth rate in each state (e.g., New York, New Jersey, Michigan, California, and Massachusetts) slowed down after the stay-at-home orders (approaching sub-exponential growth).


\section*{Discussion}
This study demonstrated a statistical relationship between two social distancing measures (travel distance and stay-at-home dwell time) and the growth rate of COVID-19 confirmed cases across U.S. states. The statistical variation of the two social distancing measures can be largely explained (with R-squared 0.60$\sim$0.69) by geographic and socioeconomic factors, including state policies, race and ethnicity, population density, age groups, and median household income (see Table S6-S9 in supplementary materials). Recent studies also identified partisan differences in Americans' response to social distancing guidelines during the COVID-19 pandemic \cite{allcott2020polarization}. One issue requires attention is that other control measures such as quarantine and enhanced personal protective procedures may also be implemented concurrently and there were no control experiments to compare such effects separately. The predictive modeling results also vary across states and the doubling time is dynamic. All these factors contribute to the endogeneity of findings \cite{jewell2020jama}. 

Great efforts have been made in scientific research communities on the study of human mobility patterns using various emerging data sources, including anonymized mobile phone call detail records \cite{gonzalez2008understanding,song2010limits,kang2012intra,gao2015spatio,yuan2016analyzing}, social media (e.g., Twitter) \cite{huang2016activity}, location-based services and mobile apps \cite{wu2014intra,shaw2016human}. During the COVID-19 pandemic, both individual-level and aggregated-level human mobility patterns have been found useful in epidemic modeling and digital contact tracing \cite{tian2020impact,ferretti2020quantifying,li2020substantial,buckee2020aggregated}. However, technical challenges (e.g., location uncertainty), socioeconomic and sampling bias \cite{zhao2016understanding,jiang2019simple,xu2018human,li2019reconstruction}, privacy and ethical concerns are raised by the national and international societies \cite{de2013unique,tsou2015research,mckenzie2019geospatial,de2018privacy}. Moving forward, research efforts will continue in exploring the balance of using such human mobility data for social goods while preserving individual rights. In summary, this study quantifies the effect of social distancing mandates on reducing the spread of COVID-19 when they are followed.



\bibliography{scibib}

\bibliographystyle{abbrv}

\section*{Acknowledgments}
We would like to thank the Descartes Labs and SafeGraph Inc. for providing the anonymous and aggregated human mobility and place visit data. We would also like to thank all individuals and organizations for collecting and updating the COVID-19 epidemiological observation data and reports. \textbf{Funding:} S.G. and J.P. acknowledge the funding support provided by the National Science Foundation (Award No. BCS-2027375). Any opinions, findings, and conclusions or recommendations expressed in this material are those of the author(s) and do not necessarily reflect the views of the National Science Foundation. \textbf{Author contributions:} Research design and conceptualization: S.G., D.D., A.K.S., J.P.; Data collection and processing: S.G., J.M.R., Y.H.K., Y.L.L.; Result analysis: S.G., J.M.R., Y.H.K., Y.L.L, A.K.S., D.D., J.F.M.R., J.P., B.S.Y.; Visualization: J.M.R., Y.H.K., Y.L.L, J.K., D.D., J.F.M.R.; Project administration: S.G.; Writing: all authors.  \textbf{Competing interests:} authors have no competing interests. \textbf{Data and materials availability:} The epidemiological data were retrieved from two sources: the COVID Atlas ( \url{https://github.com/covidatlas/coronadatascraper}) and the Department of Health Services in each state. The travel distance mobility data were provided by the Descartes Labs (\url{https://www.descarteslabs.com/mobility}). The points of interest business data with visit patterns and the home dwell time data in the United States were provided by the SafeGraph Inc. (\url{https://www.safegraph.com}).

\section*{Supplementary materials}
Materials and Methods\\
Figs. S1 to S6\\
Tables S1 to S11\\
Movie S1 to S2 \\


\pagebreak

\pagebreak

\pagebreak

\pagebreak

\begin{table}[H]
\centering
\caption{Empirical doubling time (in days) of total infected cases before and after stay-at-home orders in different states.}
\resizebox{0.65\linewidth}{!}{%
\begin{tabular}{@{}cccccc@{}}
\toprule
State Name & Before Order (Median) & After Order (Median) & Change \\ \midrule
Alabama          & 3.281                        & 6.535                               & 3.254                                \\
Alaska           & 6.856                        & 30.289                              & 23.433                               \\
Arizona          & 2.492                        & 6.815                               & 4.323                                \\
California       & 3.255                        & 5.278                               & 2.023                                \\
Colorado         & 2.648                        & 6.167                               & 3.519                                \\
Connecticut      & 1.677                        & 4.471                               & 2.794                                \\
Delaware         & 2.906                        & 4.715                               & 1.809                                \\
Florida          & 2.972                        & 9.989                               & 7.017                                \\
Georgia          & 3.484                        & 6.396                               & 2.912                                \\
Hawaii           & 1.954                        & 7.318                               & 5.364                                \\
Idaho            & 1.254                        & 4.755                               & 3.501                                \\
Illinois         & 1.940                        & 4.681                               & 2.741                                \\
Indiana          & 2.688                        & 3.655                               & 0.967                                \\
Kansas           & 2.688                        & 5.849                               & 3.161                                \\
Kentucky         & 2.541                        & 5.400                               & 2.859                                \\
Louisiana        & 2.059                        & 4.577                               & 2.518                                \\
Maine            & 3.744                        & 16.528                              & 12.784                               \\
Maryland         & 2.822                        & 4.217                               & 1.395                                \\
Massachusetts    & 3.781                        & 4.706                               & 0.925                                \\
Michigan         & 2.317                        & 4.370                               & 2.053                                \\
Minnesota        & 2.985                        & 8.673                               & 5.688                                \\
Mississippi      & 2.763                        & 9.422                               & 6.659                                \\
Montana          & 2.376                        & 8.296                               & 5.920                                \\
Nevada           & 3.735                        & 11.22                               & 7.485                                \\
New Hampshire    & 3.010                        & 5.783                               & 2.773                                \\
New Jersey       & 1.770                        & 4.216                               & 2.446                                \\
New Mexico       & 3.106                        & 5.174                               & 2.068                                \\
New York         & 1.838                        & 6.449                               & 4.611                                \\
North Carolina   & 2.652                        & 6.326                               & 3.674                                \\
Ohio             & 2.136                        & 5.279                               & 3.143                                \\
Oklahoma         & 2.409                        & 5.647                               & 3.238                                \\
Oregon           & 3.802                        & 6.731                               & 2.929                                \\
Pennsylvania     & 2.487                        & 5.767                               & 3.280                                \\
Rhode Island     & 1.943                        & 4.649                               & 2.706                                \\
South Carolina   & 2.409                        & 5.770                               & 3.361                                \\
Tennessee        & 3.338                        & 10.342                              & 7.004                                \\
Texas            & 3.432                        & 5.981                               & 2.549                                \\
Utah             & 2.515                        & 6.653                               & 4.138                                \\
Vermont          & 2.270                        & 7.100                               & 4.830                                \\
Virginia         & 3.436                        & 4.847                               & 1.411                                \\
Washington       & 4.530                        & 12.323                              & 7.793                                \\
Washington, D.C. & 3.491                        & 6.853                               & 3.362                                \\
West Virginia    & 1.038                        & 4.430                               & 3.392                                \\
Wisconsin        & 2.255                        & 6.984                               & 4.729                                \\
Wyoming          & 3.150                        & 7.879                               & 4.729        \\\bottomrule
\end{tabular}}
\label{tab:real_doublingtime_state}
\end{table}

\end{document}